\documentclass{article}

\usepackage{epsfig}
\usepackage{amssymb}

\begin{document}

\title{
Automatic Detection of Trends in Dynamical Text: An Evolutionary Approach
}
\author{Lourdes Araujo \\
{\tt lurdes@sip.ucm.es} \\
Dpto. Sistemas Inform\'aticos y programaci\'on.\\
Universidad Complutense de Madrid \and
Juan Julian Merelo\\
{\tt jmerelo@geneura.ugr.es}\\
Dpto. Arquitectura y Tecnolog\'{\i}a de Computadores,\\
ETS Ingenier\'{\i}a Inform\'atica, Universidad de Granada }

\maketitle
\begin{abstract}
This paper presents an evolutionary algorithm for modeling the arrival
dates of document streams, which is any time-stamped collection of
documents, such  as  newscasts, e-mails, IRC conversations, scientific
journals archives and weblog postings. 
This algorithm assigns frequencies (number of document arrivals per
time unit) to time intervals so that it produces an optimal fit to the data. 
The optimization is a trade off between accurately fitting the data and
avoiding too many frequency changes; 
this way the analysis is able to find fits which ignore the noise.
Classical dynamic programming algorithms are limited by memory and efficiency
requirements, which can be a problem when dealing with long streams.
This suggests to explore alternative search methods which allow for some degree of uncertainty to
achieve tractability. Experiments have shown that the designed evolutionary algorithm
is able to reach the same solution quality as those classical dynamic
programming algorithms in a shorter time.
We have also explored different probabilistic models to optimize the
fitting of the date streams, and applied these algorithms to infer
whether a new arrival increases or decreases {\em interest} in the
topic the document stream is about.  


\end{abstract}

\section{Introduction}

The analysis of information flow has become a critical task nowadays.
Constantly evolving websites, repositories of news, e-mails, chat
logs, and scientific papers
are some clear examples of streams whose interpretation highly
depends on the sequence of occurrence of the documents that constitute it.
All these examples have a temporal dimension that has to be taken into
account when analysing their content.

One of the first attempts to model the dynamic component of a stream of documents
was the Topic Detection and Tracking (TDT) research project \cite{allan98topic,allan02,papka99}, which 
among other contributions, attempted to settle the different tasks
that can be distinguished in identifying topics in documents streams.
First of all, a distinction is established between {\em topic}, i.e. a general concept, and {\em event},
i.e. an occurrence of a topic in a particular time. 
Among the distinguished tasks are the selection of {\em new events}, i.e. to identify the occurrence
of a document discussing a new event; the {\em tracking} of a detected event, i.e. 
to identify a collection of documents about the same event, and
the {\em segmentation} of the large streams of documents in substreams related
to different topics. Different kinds of techniques have been applied to perform
these TDT tasks, which involve both content similarity analysis and temporal analysis methods.

More recent papers have focused in identifying time segments in which the appearance
of documents of a particular topic can be considered relatively stable.
The frequency of occurrences can be very noisy, and this hinders the identification
of intervals of similar frequency.
Different statistical techniques\cite{kleinberg02,bingham03,girolami04} have been applied 
to analyse temporal changes in document streams.

Another approach is to  focus on studying the rise and fall
of frequencies to detect trends in streams by identifying changes in the frequencies
along given periods of time.
Charikar et al. \cite{charikar02} have proposed an algorithm for
finding the most
frequent elements in a stream, which is also adapted to find elements whose
frequencies change the most.

All these papers are reviewed by Kleinberg \cite{kleinberg05}, who also discusses
different approaches to the problem. 
 
In this work we are interested in identifying the trends in streams of documents
in a robust and efficient way. 
This is a task of great interest for newsmakers, and the society at
large: when large amounts of data are available, it is difficult to
answer the question {\em What is everybody talking about?}
Therefore, our problem has some common ground with one of those tackled in \cite{charikar02},
although in this case we focus on the relative change of 
frequencies of a single and isolated topic, not different topics as above.
Instead of adopting the approach of analysing the amount of change for fixed periods
of time, we model the flow in a segment of time for which
enough data are available, following Kleinberg's approach \cite{kleinberg02}.

The model is as follows: let us assume a stream of $N$ documents arriving along a period of time $T$,
and a set of $S$ frequencies, ranging between 0 and 1. We have
a double goal: on the one hand, we want to determine the most appropriate frequency
for each interval between one arrival and the next one; on the other hand,
we are interested in grouping together intervals with ``similar'' frequencies, i.e.
relatively stable, in order to wash out the noise. 

\begin{figure}[htbp]
\begin{center}
  \scalebox{0.8}{\epsfig{file=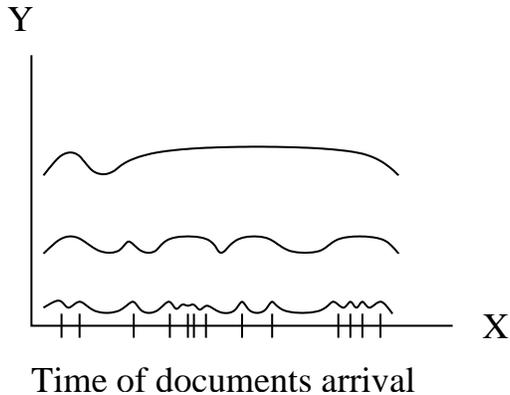,clip=}}
\caption[]{Some possibles fitting curves for a stream of documents}

\label{fajustes}
\end{center}
\end{figure}
Figure \ref{fajustes} shows different possible fitting curves for the frequencies 
of a sequence of documents which have arrived at the instants of time marked in the X axis.
If the chosen probabilistic model does not penalize changes of state, the optimum curve 
would be the one which assigns to each interval the most appropriate frequency,
which, in general, amounts to a change of state for each arrival.
In this case, the chosen fit for the arrival marked in the X axis in
Figure \ref{fajustes} would be the lowest one. 
However, we want to assign the same state to consecutive intervals with similar frequency,
ignoring in this way the changes produced by the randomness in the arrivals.
This can be done by penalizing the change of state in different manners.

Kleinberg \cite{kleinberg02}, whose method has inspired this work,
has developed a framework which formalizes these ideas. 
He proposes a probabilistic automaton to model
the frequency of appearance of documents in different intervals.
The state of the automaton at a particular time determines
the expected frequency of document occurrences,
while transitions between states are probabilistically modeled \cite{elwalid93}.
A burst of documents for a topic can therefore be identified by the
period in which the automaton has stayed in a high frequency state.
Given a stream of documents related to a particular topic, the Viterbi
dynamic programming algorithm \cite{Forney73}, can be applied to determine the
sequence of automaton states which optimizes the measure defined by 
the chosen probabilistic model.
This algorithm was initially designed to find
the most probable path in a Markov chain which produces a given sequence of tags,
and it is now widely applied to a large range of problems \cite{rabiner90}.

The streams of document dates to model can be obtained from very different
environments, though in many cases they are long sequences of documents appearing along unlimited
periods of time, such as chats lots, e-mails and news stories.
Since a dynamic programming algorithm is limited by memory and efficiency
constraints, as the length of the stream or the number of automaton states grows,
it makes sense to explore alternative search methods where a degree of uncertainty 
is allowed in order to achieve tractability.
We have designed an evolutionary algorithm to perform the search
for the optimal sequence of states according to the selected model.
Evolutionary algorithms (EAs) do not guarantee  reaching the optimum solution, but
always produce a reasonably good approximation, according to the resources 
assigned (time and memory).  
On the one hand, an approximate solution is always better than no solution at all;
on the other hand, one should bear in mind that the probabilistic model used to 
penalize the state transitions and deal with noisy events,
is just an approximation itself, which can be taken in different ways leading to
different results, as the experiments presented later on will show.
What is really interesting for most applications, such as the detection of
trends in streams of documents, is the detection of significant changes of state,
i.e. the detection of clear changes in the average intensity of document arrivals,
and not obtaining a very precise optimum numerical result for the cost function.
Because of these reasons, this work investigates the application of evolutionary 
algorithms to the problem.

Besides using an alternative algorithm for fitting frequencies, in
this paper we have also studied alternative penalization functions to the one proposed by
Kleinberg \cite{kleinberg02} for modeling the change of state when new document arrive. 
We have built a number of artificial streams, for which we know the best fit, in order
to be able to compare the results obtained with the different cost functions.


The rest of the paper proceeds as follows:
section 2 describes the model proposed by Kleinberg, which has
inspired this work, and the different variant we are evaluating;
section 3 is devoted to describe an evolutionary algorithm used to find
the optimal fit of frequency assignments;
section 4 presents and discusses the experimental results,
and section 5 draws the main conclusions from this work, and discusses
future lines of research.

\section{Models for the Problem}
\label{smodels}

Our intention in this paper is to devise a model which generates a
sequence of events and set its 
parameters in such a way that the stream under study might have been generated
with this model with a high probability.
In order to model the arrival of documents in a stream we are going to use a finite
state automaton (FSA) as in Kleinberg's approach \cite{kleinberg02},
inspired in turn in models for network traffic in queuing theory \cite{elwalid93}
and on Hidden Markov Models \cite{rabiner90}.
According to this approach, a source of traffic emits documents at a rate
which depends on the state of the FSA at a given point in time.
A traffic burst begins with a transition from a state of lower rate of emission 
to one of higher rate.
Kleinberg chooses an exponential density function $f(x) = \alpha e^{-\alpha x}, \alpha > 0$
to model the probability density of waiting times between arrivals.
In the simplest case, we can distinguish between only two different states, with high and
low frequency respectively. In state $q_0$ the automaton emits documents at a low rate, which 
gives rise to a probability density for the intervals $f_0(x) = {\alpha}_0 e^{-{\alpha}_0 x}$,
while state $q_1$ has a higher emission rate, which gives rise to a probability density
for the gaps given by $f_1(x) = {\alpha}_1 e^{-{\alpha}_1 x}$, with ${\alpha}_1 > {\alpha}_0$.
Now the question is how to model the probability $p$ with which the automaton changes its
state between the emission of two consecutive documents. It is assumed that $p$ is
independent of previous emissions and state transitions.
Assuming this probability distribution, we can compute the probability of a sequence 
$q = \{q_1,\cdots,q_n\}$ of state transitions conditioned to the sequence $x = \{x_1,\cdots,x_n\}$
of gaps observed between the $n+1$ documents arrived in the stream.
The state sequence which maximizes this probability $P(q|x)$ is the one which minimizes
the cost function $c(q|x) = - \mbox{ln} P(q|x)$. Applying this condition, Kleinberg obtains the formula
\begin{equation}
  c({\bf q}|{\bf x}) = b ~\mbox{ln}~ (\frac{1 - p}{p}) + \sum_{t=1}^n - \mbox{ln}~ f_{i_t}(x_t)
\label{model_2est}
\end{equation}
where $b$ is the number of state transitions done to emit the sequence, i.e. the number
of times in which $q(t) \neq q(t+1)$. In formula (\ref{model_2est}),
we can observe that the fewer state transitions, the smaller the first term,
while the better the state sequence fits the observed sequence of gaps $x$, the smaller the second term.
Therefore, it is expected that the optimum sequence
of states fits well the gap sequence with as few 
changes in the size of the gaps as possible, depending this inertia on the parameter $b$.

Afterwards, Kleinberg extends this simple two-state model to one of infinite states,
providing a different one for each possible intensity of emission.
Kleinberg proposes an automaton with an initial state $q_0$ whose corresponding
density function ${\alpha}_0 e^{-{\alpha}_0 x}$ is assigned an emission rate
${\alpha}_0 = n/T$, where $n$ is the number of gaps between documents emissions
and $T$ is the total length of the considered period of time; 
i.e., ${\alpha}_0$ corresponds to a perfectly uniform event emission.
For the remaining states $q_i, i >0$, the assigned emission rate is ${\alpha}_i = {\alpha}_0 s^i$,
where $s > 1$ is a scaling parameter, i.e. the smaller the gap, the greater the intensity.
Then, by analogy with the two-state model, the cost function which the 
selected sequence of states must minimize, is
\begin{equation}
c({\bf q}|{\bf x}) = \sum_{t=0}^{n-1} \tau(i_t, i_{t+1})  + \sum_{t=1}^n - \mbox{ln}~ f_{i_t}(x_t)
\end{equation}
where $\tau(q_i, q_j)$ represents the cost of a state transition from the state $q_i$ at 
a given time $t$ to the state $q_j$ at time $t+1$.
Kleinberg, which considers that the selection of $\tau(q_i, q_j)$ is very flexible, chooses
$\tau(q_i, q_j)$ in such a way that the cost of changing from a lower intensity state
to a higher intensity one is proportional to the number of involved states, while there
is no cost for changing for higher to lower intensity states. 
Specifically, the cost associated to change from state $q_i$ to $q_j$, where $j > i$, is
defined as $(j - i) \gamma \mbox{ln} n$, $\gamma$ being a parameter of the model.
$\mathcal{A}_{s,\gamma}^*$ denotes the automaton of infinite states with parameters $s$ and $\gamma$.
The parameter $s$ controls the scale for the rate values of the states, while $\gamma$, which
Kleinberg sets to 1 in his experiments, controls the resistance to changing state.

Kleinberg shows that computing an optimal state sequence in an automata $\mathcal{A}_{s,\gamma}^*$
with infinite states is equivalent to compute $q$ in one of its finite restrictions
$\mathcal{A}_{s,\gamma}^k$, obtained by deleting from the automaton all states but the first $k$ of them.
This result allows establishing algorithms to compute the sequence of states for the
minimum cost. For this purpose, Kleinberg adopts the standard dynamic programming
algorithm used for hidden Markov Models.

However, this penalty function is not the only possible one. We have
studied other penalization functions. On the one hand, we have tested 
cost functions which also penalize the change from a state with high intensity to a state of low
intensity. On the other hand, we have searched for measures which do not
depend on the number of states. We have chosen the following set of measures:

\begin{itemize}
\item
${\tau}_a$, defined as  
$
\left\{ \begin{array}{l}
          (j - i) {\gamma} \mbox{ln} n ~~\mbox{if}~~ j > i\\
          0,~~ \mbox{if}~~ i \leq j 
       \end{array}
\right.
$, the one proposed by Kleinberg, where $n$ stands for
the number of documents in the stream.
\item
${\tau}_b$, defined as  
$|j - i| {\gamma} \mbox{ln} n$, similar to ${\tau}_a$ but now there is also a 
cost by changing from a higher intensity state to a lower one.
\item
${\tau}_c$, defined as  
$
\left\{ \begin{array}{l}
          \log(j - i) {\gamma} ~~\mbox{if}~~ j > i\\
          0,~~ \mbox{if}~~ i \leq j 
       \end{array}
\right.
$, other penalization function with zero penalty for high-to-low intensity changes.
\item
${\tau}_d$, defined as  
$\log(|j - i|) {\gamma} $, the counterpart of ${\tau}_c$ which also penalizes 
high-to-low intensity changes.
\item
${\tau}_e$, defined as  
$
\left\{ \begin{array}{l}
          \sqrt{(j - i)} {\gamma} ~~\mbox{if}~~ j > i\\
          0,~~ \mbox{if}~~ i \leq j 
       \end{array}
\right.
$, which has zero penalty for high-to-low intensity changes.
\item
${\tau}_f$, defined as  
$\sqrt{|j - i|} {\gamma} $, the counterpart of ${\tau}_e$.
\item
${\tau}_g$, defined as  
$
\left\{ \begin{array}{l}
          (j - i) {\gamma} / \mbox{ln} E ~~\mbox{if}~~ j > i\\
          0,~~ \mbox{if}~~ i \leq j 
       \end{array}
\right.
$, introduced to avoid the dependency with the number of states.
$E$ is the total number of automaton states.
\item
${\tau}_h$, defined as  
$|j - i| {\gamma} / \mbox{ln} E$, the counterpart of ${\tau}_g$ which penalizes any change of state.
\end{itemize}
The results obtained with the different functions have been compared in the
section devoted to the experiments (section \ref{sec:exp}).

\section{The Search Algorithm: Evolutionary Algorithm}

Systems based on evolutionary algorithms \cite{michalewicz04} maintain a population of potential
solutions to the problem and apply some selection process based 
on the quality or  {\em fitness} of individuals, as natural selection does. 
The population is renewed by replacing individuals with those
obtained by applying ``genetic'' operators to selected individuals.
The most usual ``genetic'' operators are {\em crossover} and {\em mutation}.
Crossover obtains new individuals by mixing, often in some problem dependent way,
two individuals, called parents. Mutation produces a new individual by
performing some kind of random change on an individual.
The production of new generations continues  until  resources  are
exhausted or until some individual in the population is fit enough.
Evolutionary algorithms have proved very useful as search and
optimization methods, and have previously been applied
to different issues of natural language processing \cite{Kool00,Araujo04_IEEE}, 
such as text categorization, inference of context-free grammars, tagging and parsing.

In our case, individuals represent sequences of state transitions in the automaton.
The fitness of individuals is the cost function associated to the sequence of state
transitions, and depends on the chosen probabilistic model.

\subsection{Individual representation}

Let $E$ the number of states chosen for the automaton. Let us assume a stream 
of $n+1$ documents arriving along a period of time $T$.
Then, the individuals of our evolutionary algorithm could be represented
as the list of automaton states corresponding to each arrival of a
document. Accordingly, an individual would be a list of $n$ genes $g_i$, where
$ g_i \in \{0,\cdots,E\}$ is the state $q$ in which the automaton is after the arrival of 
document $i$.

\begin{center}
\begin{tabular}{|c|c|c|c|c|}
\hline
$q(t_1)$ & $q(t_2)$ & $\cdots$ & $q(t_n)$ \\
\hline
\end{tabular}
\end{center}
However, the arrival of a new document does not produce a state transition in
many cases. Therefore, the sequence of transitions can be represented in a more compact
manner. Thus, an individual is a variable length list, in which each position, 
or ``gene'', represents the arrival of a subsequence of documents which do not
lead to a change of state. Each gene is composed of an automaton state and of an identifier
of the first and the last document in the subsequence.
Therefore, the individuals of our evolutionary algorithm could be represented
as the list of state transitions in the automaton caused by the arrival of the
documents.

\begin{center}
\begin{tabular}{|c|c|c|c|c|}
   $g_1$                  & $g_2$                                    & $\cdots$ & $g_f$\\
\hline
$q(t_1), (t_1, t_{k_1})$  & $q(t_{k_1} + 1), (t_{{k_1}+1}, t_{k_2})$ & $\cdots$ & $q(t_{f_1}), (t_{{f-1}+1},t_n)$ \\
\hline
\end{tabular}
\end{center}

\subsection{The Fitness Function}

For the fitness function we take, quite naturally, the
cost function which defines the chosen statistical model.
Thus the goal of our evolutionary algorithm is to find the sequence of state
transitions {\bf q} = $(q_{i_1}, \cdots, q_{i_n})$ which minimizes the function
\[
c({\bf q}|{\bf x}) = \sum_{t=0}^{n-1} \tau(i_t, i_{t+1}) + \sum_{t=1}^n - \mbox{ln}~ {\alpha}_i e^{-{\alpha}_i x_t},
\]
with the penalty cost $\tau(i, j)$ and the state parameter, ${\alpha}_i$, of the
chosen statistical model.
In order to compute this function, the implicit automaton underlying the model must be
completely defined, i.e. we have to assign values to each ${\alpha}_i$.
We assume that the number of automaton states has previously been fixed to $E$.
Following Kleinberg's approach, we establish a {\em uniform state} $q_0$, with a
document arrival rate ${\alpha}_0 = n/T$, which corresponds 
to uniform document arrivals.
For the remaining states, $q_i, i > 0$, the arrival rate is ${\alpha}_i = {\alpha}_0 s^i$, 
where $s > 1$ is a scaling parameter, i.e. the arrival intensity increases geometrically
with $i$.
 
If we knew the value for the rate ${\alpha}_i$ of a particular state, we could obtain
the value for $s$ from:
\[
{\alpha}_i = \frac{n}{T} s^i.
\]
By assigning to ${\alpha}_E$, the maximum arrival rate in the automaton,
a particular value, such as $1$, we obtain a value for $s$:

%

\[
s = \exp\{(\mbox{ln}~{\alpha}_E - \mbox{ln}~n + \mbox{ln}~T)/E\}
\]

Depending on the form of the selected cost function $\tau(i,j)$ for a state transition from state
$q_i$ to $q_j$, further parameters must be fixed.
For example, if we use the cost function ${\tau}_a$ of section \ref{smodels}, the one used by Kleinberg,
we must give a value to the parameter $\gamma$, on which the inertia of the
automata to change its state depends. In his experiments, Kleinberg chose $\gamma = 1$,
as we also do.

\subsection{Initial Population}

The first step of an evolutionary algorithm is the creation of a population
of individuals, or potential solutions to the problem. In our case, these individuals
represent sequences of state transitions randomly generated.
The simplest way of creating one such sequence is to choose a few documents at random 
and use them to split the whole document stream into intervals, each of which is assigned a random state.
However, some preliminary experiments we have performed have shown that 
such a simple strategy gives rise to a 

search space that is too large for the algorithm to be efficient. 
Accordingly, we choose for a state transition only those documents for which the 
gaps with the previous document and with the following one are sufficiently 
different (the size of one at least 50 \% longer than the other).
The interval before the first transition is assigned a random state, and this state is
increased or decreased in successive intervals according to whether the gaps on the left
and right of the partition points increase or decrease in length, respectively.
The size of this change is randomly chosen.

\subsection{Crossover Operator}

We have implemented the classic one point
crossover, which creates two offspring by combining two individuals in such a way that
the first part of one parent up to a crossover point is combined with
the second part of the  
other parent and vice versa.
Afterwards, the best offspring substitutes the worst parent. This is a
steady state, elitist strategy.

The steps to apply this operator are the following:
\begin{itemize}
\item
In order to select the crossover point, we randomly select one of the dates
of document arrival in the stream. Then, we search in both parents the gene which
contains this document.
\item
Then, the genes on the right-hand side of the selected gene in both parents are exchanged.
\item
For the gene containing the crossover point, we must decide if the substream
of documents ---which, in general, is different in both parents--- is going to be joined or split.
Experiments have shown that taking this decision at random produces bad results.
Accordingly, if the gaps to the left and to the right of the crossover point are
comparable, the substream is assigned a single gene whose state is
randomly selected from one of the parents. Otherwise, the substream is split
at the crossover point in two genes, each taking the state from one parent.
\end{itemize}

\subsection{Mutation Operator}

The mutation operator is applied to every individual of the population 
with a probability given by the mutation rate.
Different variants of mutation have been implemented, selecting at random the one
to apply in each case: 
\begin{itemize}
\item
One of these mutation operators amounts to choosing a gene at
random and randomly increment or decrement its state by one unit.
\item
Other mutation operator joins two consecutive genes to produce a single one.
The state of the new gene is randomly taken from one of the original genes.
\item
The last mutation operator splits a gene in two; each one is assigned
a different state: one of them is given the state of the original gene and
the other one is given the previous state plus or minus one (plus if the gap
on the left of the partition point is longer than the one on the right, and minus
otherwise). This operator is only applied if the gaps in both sides of the 
partition are different. 
\end{itemize}

\section{Experimental Results}
\label{sec:exp}
The first step in our experiments has been to investigate which cost function
works better. To do this, we have generated
artificial streams with different features, for which we
know the probability distribution of the document arrivals.
The results of an EA are usually sensitive to settings such
as the population size, the number of generations and the rates of application
of the crossover and mutation operators.
Because of this, we have investigated
the range of values for these parameters which provide best results.
We have also compared the fits obtained with a dynamic programming algorithm with
the EA.
Finally, we have also tested our system in some real world streams.

\subsection{Evaluating Cost Functions}

In principle, it is not a trivial task to compare the results obtained
using different cost functions.  
It is pointless to compare the numerical value provided by each of them for the cost, 
because, in general, they are going to be different even for the same fitting curve. 
The best way to perform the comparison is to apply the different measures to a stream
for which we know the emission frequencies that generated it. In this case,
we can compare the correct fitting curve, with the curves obtained with the different
cost functions.
Because  real world streams present too much noise to intuitively determine
the best fit, we have applied this method to two kinds of artificial
streams, 
composed of a number of intervals spanning the total time-length.
These intervals have been designed by hand to present different degrees of
difficulty for the algorithm.
The program which generates streams of the first kind reads from an input file the 
number of documents arriving within each interval and the gap between every two
consecutive ones, which is constant along the interval.

\begin{figure}[tbp]
\begin{center}
  \scalebox{0.3}{\epsfig{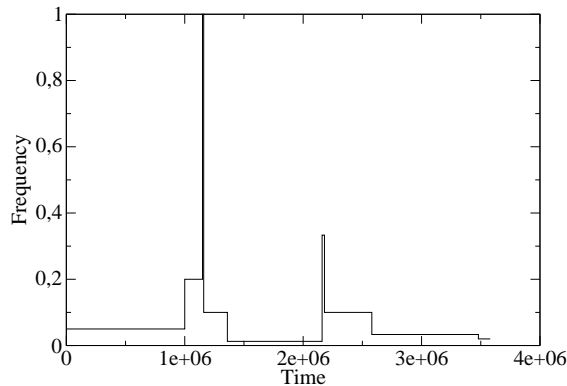}}\\
\caption[]{Artificial stream generated to evaluate cost functions.}
\label{fartificial_c}
\end{center}
\end{figure}

Intervals correspond to steps in the frequency-time representation.
Figure \ref{fartificial_c} represents a stream of this type, which is composed of
220000 dates and presents ascending and descending steps.

\begin{figure}[tbp]
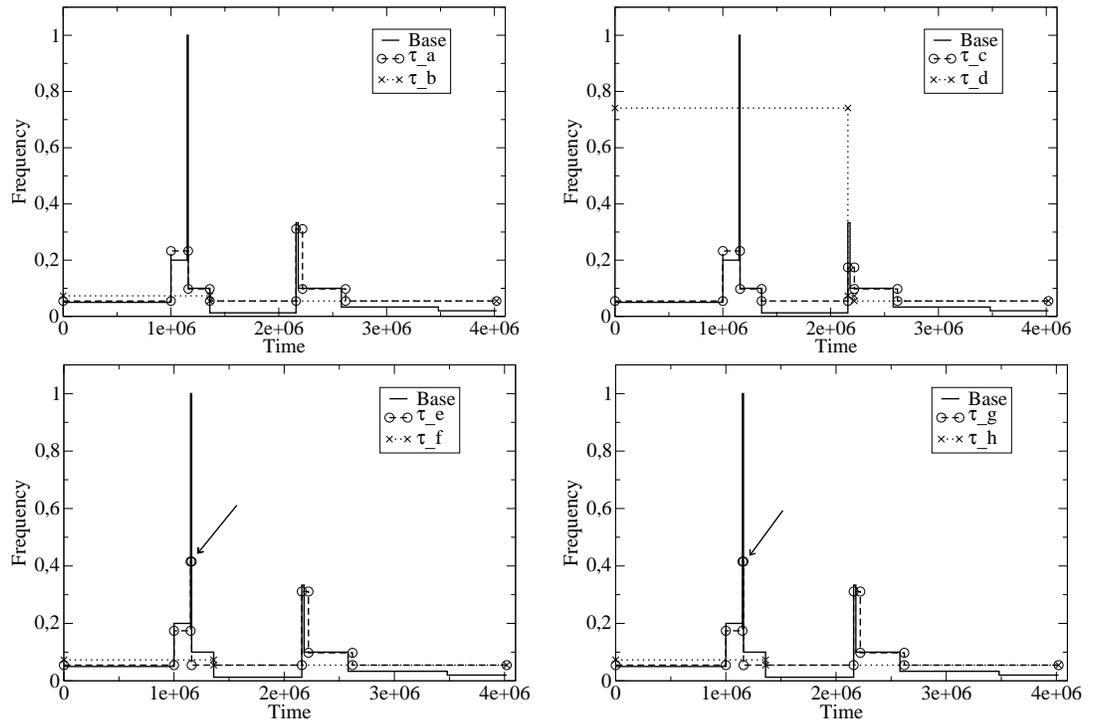

\begin{center}
  \begin{tabular}{ll}
  \scalebox{0.28}{\epsfig{file=dibujos/cost3_ab.eps,clip=}} &
  \scalebox{0.28}{\epsfig{file=dibujos/cost3_cd.eps,clip=}}\\
  \scalebox{0.28}{\epsfig{file=dibujos/cost3_ef.eps,clip=}} &
  \scalebox{0.28}{\epsfig{file=dibujos/cost3_gh.eps,clip=}}\\
  \end{tabular}
\caption[]{Fitting curves obtained with different cost functions for the stream
of Figure \ref{fartificial_c}. Clockwise from the top-left corner,
${\tau}_a$ and ${\tau}_b$, ${\tau}_c$ and ${\tau}_d$, ${\tau}_g$ and ${\tau}_h$, ${\tau}_e$ and ${\tau}_f$.}
\label{fcost3}
\end{center}
\end{figure}

Figure \ref{fcost3} shows the fitting curves obtained by applying the Viterbi
algorithm with the different cost functions
for the stream of Figure \ref{fartificial_c}. We can observe 
that the best fittings, i.e. closest to the {\em base} curve 
of frequencies used for generating the stream, are obtained with functions ${\tau}_e$ and ${\tau}_g$. 
Notice that these functions are able to detect the narrow peak in the stream
(marked with an arrow in the figure).
However, we choose ${\tau}_g$ because it is normalized with respect to the
number of states, which 
makes its behavior more insensitive to this parameter.
Measures ${\tau}_b$, ${\tau}_d$, ${\tau}_f$ and ${\tau}_h$ whose changes 
from a higher frequency state to a lower one
are also penalized, produce fitting curves that are much too flat.

\begin{figure}[tbp]
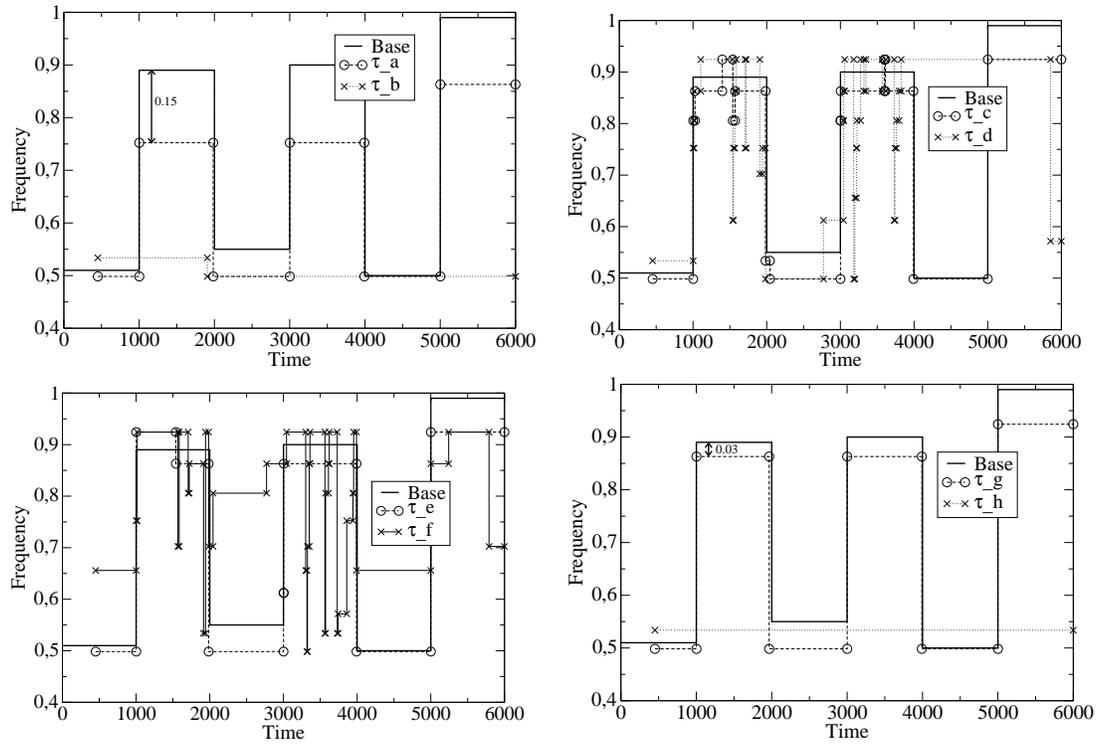

\begin{center}
  \begin{tabular}{ll}
  \scalebox{0.28}{\epsfig{file=dibujos/alea1_ab.eps,clip=}} &
  \scalebox{0.28}{\epsfig{file=dibujos/alea1_cd.eps,clip=}}\\
  \scalebox{0.28}{\epsfig{file=dibujos/alea1_ef.eps,clip=}} &
  \scalebox{0.28}{\epsfig{file=dibujos/alea1_gh.eps,clip=}}\\
  \end{tabular}
\caption[]{Fitting curves obtained with different cost functions for the stream
generated with the probability distribution appearing in Table \ref{talea1}.}
\label{ftramo_alea}
\end{center}
\end{figure}

We have also used for the evaluation another kind of stream with random gaps.
To generate it, every time unit we decide that a document arrives or not with a given
probability (its frequency).
Figure \ref{ftramo_alea} shows the fits obtained by applying the Viterbi algorithm
to an automaton of 10 states for a stream generated according to the
probability distribution of Table \ref{talea1}.

The first row of this table indicates the initial dates of each segment, the second one the last
date, and the third one is the average frequency at which documents are emitted.
Gaps between documents in a segment are thus random.
In Figure \ref{ftramo_alea} the most accurate and smoothest fit is again obtained with 
the cost function ${\tau}_g$. This function and ${\tau}_a$ are the ones which
most accurately reproduce the frequency changes. However, ${\tau}_g$
achieves a better fit of the frequencies.
In this case, cost functions ${\tau}_b$, ${\tau}_d$, ${\tau}_f$ and ${\tau}_h$, 
are too sensitive to the noise, overfitting the arrival time curve. 
\begin{table}[htb]
\begin{center}
\begin{tabular}{|l|l|l|l|l|l|l|}
\hline
Ini    &    0    &   1001  &  2001  & 3001   & 4001   & 5001\\
\hline
End    &  1000   &   2000  &  3000  & 4000   & 5000   & 6000\\
\hline
Freq.  & 0.002   &  0.89   &  0.004 & 0.9    & 0.001  & 0.99\\
\hline
\end{tabular}
\caption{Probability distribution used to generate a stream with random gaps.
}
\label{talea1}
\end{center}
\end{table}

As a conclusion, we have used ${\tau}_g$ instead of the cost function
proposed by Kleinberg in his paper, since it provides a better fit for
known document streams. This function avoids the dependency with the
number of states that was featured by Kleinberg's proposed one; the
main difference is that while Kleinberg's function (${\tau}_a$ here) uses
as a scale factor the number of documents, ours (${\tau}_g$ here) scales by
the number of states. As can be seen in figures \ref{fcost3} and
\ref{ftramo_alea}, the difference is bigger in the case of a
randomly generated stream (\ref{ftramo_alea}) than in the case of a 
stream with uniform gaps (\ref{fcost3}), which probably means that, in
the general case, ${\tau}_g$ will yield better results.

\subsection{Evolutionary Algorithm Parameters}

In EAs there is always a trade-off between two fundamental factors in
the evolutionary process: population diversity and selective pressure.
An increase in selective pressure will induce a decrease in
population diversity, leading to a premature convergence of the EA, while
a weak selective pressure can made the search ineffective.
The relation between these factors is mainly determined by the population size,
and the rates of application of crossover and mutation.
A too small population size lacks the required diversity and the algorithm
usually converges to a bad result. However, much too large populations require a
long time to converge to a solution. 

Crossover and mutation rates, 
also critical for the effectiveness of the algorithm, must be correlated with 
the population size to provide the EA with a suitable population diversity.
The most appropriate values for these parameters are strongly problem-dependent.
\begin{figure}[tbp]
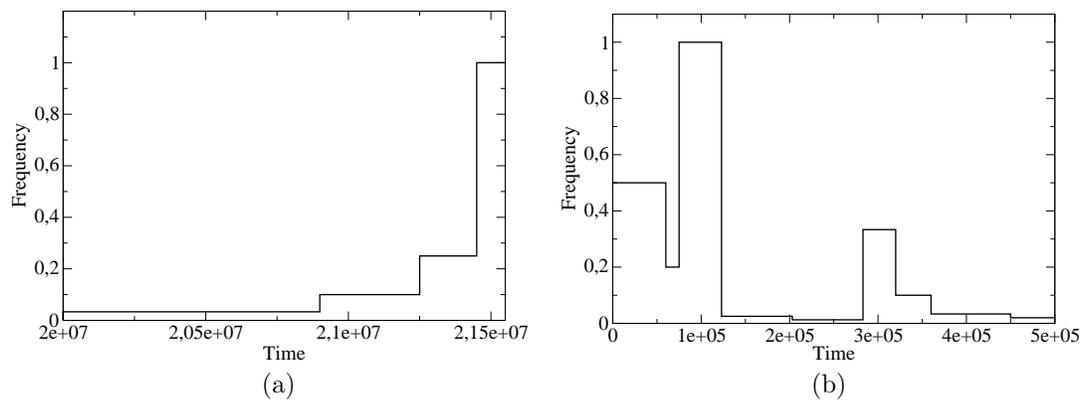

\begin{center}
 \begin{tabular}{ll}
  \scalebox{0.28}{\epsfig{file=dibujos/art_param4.eps,clip=}} &
  \scalebox{0.28}{\epsfig{file=dibujos/art_param5.eps,clip=}}\\
  \hspace*{3.22cm} (a) & \hspace*{3.22cm} (b)\\
 \end{tabular}
\caption[]{Additional artificial streams generated to evaluate the EA.}
\label{fartificial}
\end{center}
\end{figure}
We have used the artificial streams (a) and (b) depicted in Figure \ref{fartificial} and
the stream of Figure \ref{fartificial_c}, from the previous subsection, to study 
the best parameter settings for the EA. 
The first one (a) is characterized by ascending steps. Its length is 220000 dates of document arrivals.
Stream (b) presents both, ascending and descending steps and has a length of 129000 dates.
Stream (c), also presents both kinds of steps, but in this case, the length of
the steps is very different.

\begin{figure}[htbp]
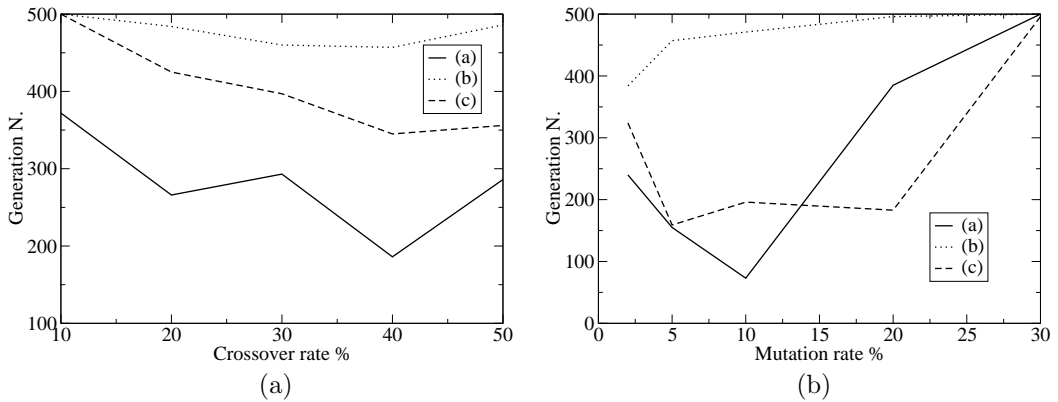

\begin{center}
 \begin{tabular}{ll}
  \scalebox{0.28}{\epsfig{file=dibujos/cruce_param.eps,clip=}} &
  \scalebox{0.28}{\epsfig{file=dibujos/mut_param.eps,clip=}}\\
  \hspace*{3.22cm} (a) & \hspace*{3.22cm} (b)\\
 \end{tabular}
\end{center}
\caption[]{Number of generations required to reach convergence (as
defined in the text)
with different rates of crossover and mutation. 
Results correspond to the streams in Figure \ref{fartificial} (a and b) and to
the stream of Figure \ref{fartificial_c} (c).}
\label{foperadores}
\end{figure}

Figures \ref{foperadores}(a) and \ref{foperadores}(b) show the number of iterations required
for the EA to reach convergence for the streams (a), (b) of Figure \ref{fartificial},
and stream (c) from Figure \ref{fartificial_c},
with different rates of application of crossover and mutation, respectively.
Convergence is achieved if the difference between the average fitness value of successive
generations lies below a threshold for a number of generations.

We can observe that intermediate values of these parameters, such as a crossover rate of 40 \% 
and a mutation rate of \%10, are enough to quickly reach convergence,
which makes the EA perform efficiently.

\begin{figure}[htbp]
\begin{center}
 \begin{tabular}{ll}
  \scalebox{0.28}{\epsfig{file=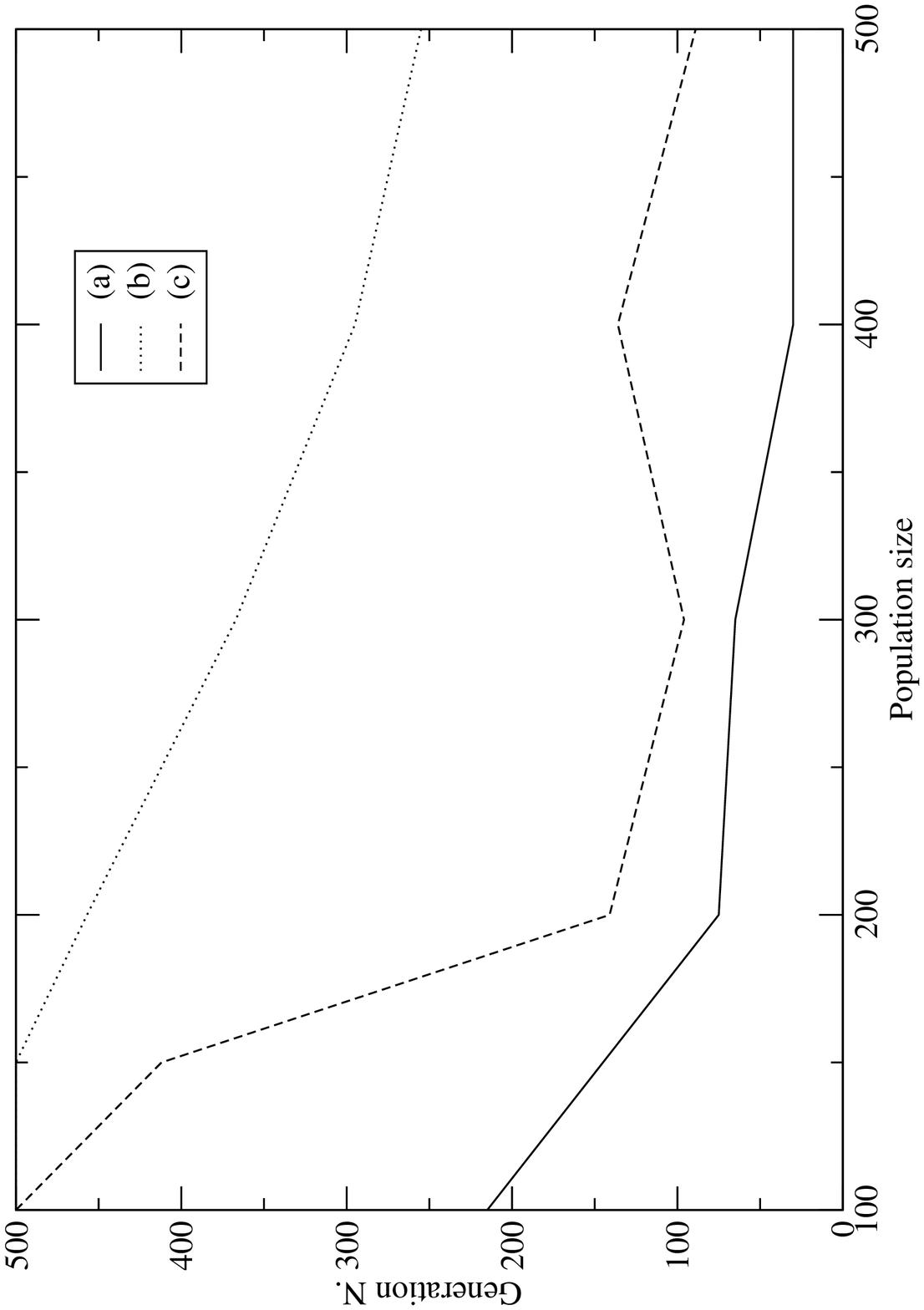,clip=}} &
  \scalebox{0.28}{\epsfig{file=dibujos/ites_param.eps,clip=}}\\
  \hspace*{3.22cm} (a) & \hspace*{3.22cm} (b)\\
 \end{tabular}
\end{center}
\caption[]{Number of generations requires to reach convergence with different population sizes (a)
and evolution of the cost function with the number of generation (b).
Results correspond to the streams in Figure \ref{fartificial} (a and b) and to
the stream of Figure \ref{fartificial_c} (c).}
\label{fpopu_ites}
\end{figure}

Figure \ref{fpopu_ites}(a) presents the number of iterations required to reach the optimum value
for the stream depicted in Figure \ref{fartificial} using different population sizes.
We can observe that a population size of 200 individuals is enough.
Larger sizes, although are also valid, increase the execution time.
Figure \ref{fpopu_ites}(b) shows the value of the cost function as the number of iterations 
increases. This chart shows that convergence can be reached very quickly.

\subsection{Comparing the EA with a classic algorithm}
In order to test the advantages of using an EA for the considered problem, we have compared
its results with those obtained with the Viterbi dynamic programming algorithm.
Tables \ref{ttiempos_a}, \ref{ttiempos_b} and \ref{ttiempos_c} show the values obtained 
for the cost function and the execution 
time when using a dynamic programming algorithm and the EA (best result of five runs,
average and standard deviation). 

\begin{table}[htbp]
\begin{center}
\begin{tabular}{|r||c|c||c|c|c|c|}
\hline\hline
State n.  & \multicolumn{2}{c||}{Viterbi} & \multicolumn{4}{c|}{Evo. Alg} \\
\hline
          & Ex. time & Cost               & Ex. time & Cost    & Av. Cost & Std. dev. \\
\hline
5         & 2394.44  & 734721             &  614.83  & 734721  & 735242.6 & 824.72     \\
\hline
10        & 4789.13  & 704021             & 1925.61  & 704021  & 707342.2 & 3081.98    \\
\hline
15        & 7188.22  & 699429             & 2688.49  & 699429  & 702288.4 & 2989.61    \\
\hline
20        & 9588.8   & 696703             & 2745.35  & 696703  & 696703   & 0   \\
\hline
25        & 12001.2  & 696210             & 1220.35  & 696210  & 696210   & 0   \\
\hline
\end{tabular}

\caption{Execution time in seconds and value of the cost function for the 
stream of Figure \ref{fartificial} (a) when applying the Viterbi and the
Evolutionary algorithms to find the optimal sequence of states in automata with
different number of states.
The cost function used has been ${\tau}_g$. The EA has been run with a population size
of 200 individuals, a maximum number of 200 iterations, a crossover rate of 40 \% and
a mutation rate of 5\%. 
The values presented for the EA are the best cost (and the running
time associated with it), the average and the standard deviation of five runs.
}
\label{ttiempos_a}
\end{center}
\end{table}

\begin{table}[htbp]
\begin{center}
\begin{tabular}{|r||c|c||c|c|c|c|}
\hline\hline
State n.  & \multicolumn{2}{c||}{Viterbi} & \multicolumn{4}{c|}{Evo. Alg.} \\
\hline
          & Ex. time & Cost               & Ex. time & Cost    & Av. Cost & Std. dev. \\
\hline
5         & 792.08   & 279464             & 632.63   & 279464  & 280897.6 & 839.70    \\
\hline
10        & 1547.4   & 277796             & 1515.57  & 277893  & 278785.4 & 899.03    \\
\hline
15        & 2319.36  & 277402             & 1678.61  & 277712  & 279385.6 & 980.11    \\
\hline
20        & 3117.28  & 277306             & 2182.12  & 277528  & 278980.4 & 1114.91   \\
\hline
25        & 3835.37  & 277260             & 2033.81  & 277270  & 279472.6 & 1116.03   \\
\hline
\end{tabular}
\caption{Execution time in seconds and value of the cost function for the 
stream of Figure \ref{fartificial} (b) when applying the Viterbi and the
Evolutionary algorithms to find the optimal sequence of states in automata with
different number of states.
The cost function employed has been ${\tau}_g$. The EA has been run with a population size
of 200 individuals, a maximum number of 200 iterations, a crossover rate of 40 \% and
a mutation rate of 5\%. 
The values presented for the EA are the best cost (and the running
time associated with it), the average and the standard deviation of five runs.
}
\label{ttiempos_b}
\end{center}
\end{table}

\begin{table}[htbp]
\begin{center}
\begin{tabular}{|r||c|c||c|c|c|c|}
\hline\hline
State n.  & \multicolumn{2}{c||}{Viterbi} & \multicolumn{4}{c|}{Evo. Alg.} \\
\hline
          & Ex. time & Cost               & Ex. time & Cost    & Av. Cost & Std. dev. \\
\hline
5         & 2377.35  & 795853             & 1719.27  & 795853  & 799612.8 & 3098.99   \\
\hline
10        & 4781.91  & 794821             & 3353.18  & 794821  & 795938.2 & 1117.64   \\
\hline
15        & 5666.24  & 794381             & 3984.38  & 794381  & 798642.0 & 3007.618  \\
\hline
20        & 9559.78  & 794256             & 5804.92  & 794256  & 797285.0 & 3023.408  \\
\hline
25        & 11975.5  & 794199             & 5550.31  & 795013  & 798818.8 & 3448.08   \\
\hline
\end{tabular}
\caption{Execution time in seconds and value of the cost function for the 
stream of Figure \ref{fartificial_c} when applying the Viterbi and the
Evolutionary algorithms to find the optimal sequence of states in automata with
different number of states.
The cost function employed has been ${\tau}_g$. The EA has been run with a population size
of 200 individuals, a maximum number of 200 iterations, a crossover rate of 40 \% and
a mutation rate of 5\%. 
The values presented for the EA are the best cost (and the running
time associated with it), the average and the standard deviation of five runs.
}
\label{ttiempos_c}
\end{center}
\end{table}

\begin{figure}[tbp]
\begin{center}
  \scalebox{0.3}{\epsfig{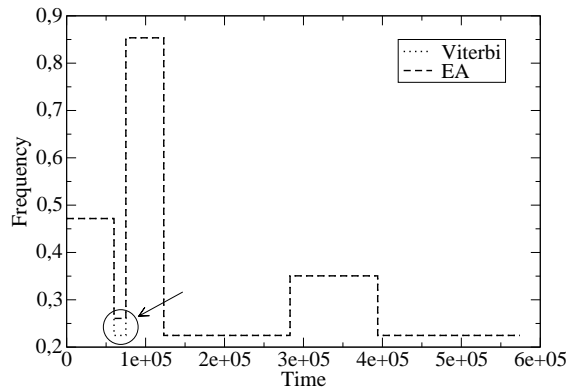}}\\
\caption[]{Steps of the fitting curve obtained with the Viterbi algorithm
and the Evolutionary algorithm
for the stream (b) from Figure \ref{fartificial} with a 10 state automaton. 
The EA is run on a population size of 200 individuals for a maximum number
of 200 generations, with a crossover rate of 40 \% and a mutation rate of 5 \%.
}
\label{ffit_example}
\end{center}
\end{figure}

It can observed  that the EA always yields the shortest running time. In some cases
the value obtained for the cost function with both algorithms and the number 
of states in the automaton is the same. There are other cases in which the
value provided by Viterbi is slightly better. However, in all these cases the fitting curve 
obtained with both algorithms is the same and the only difference is the absolute state number 
assigned to different steps; but the relative change of
state, and therefore of frequency, is maintained.
For example, in the case of the stream (b), Figure \ref{ffit_example} shows the
results obtained with an automaton of 10 states, with the Viterbi algorithm
and the EA. 
We can observe that the number of steps found in each curve is the same,
the size of each step and the document of the beginning and the end of each step is also the same in
both algorithms, and the relative change of state is also maintained.
The only change, marked in the figure, is the absolute state assigned to the interval between 
the documents 60000 and 75000.
Therefore, both solutions are equally useful in order to detect the
burst of interest in the stream and to estimate the future trend. 
Furthermore, the values obtained for the average cost and standard deviation, which
is below 0.5\% in all cases, show that the EA is highly robust.

\subsection{Evaluating Real World Document Streams}
\begin{figure}[htbp]
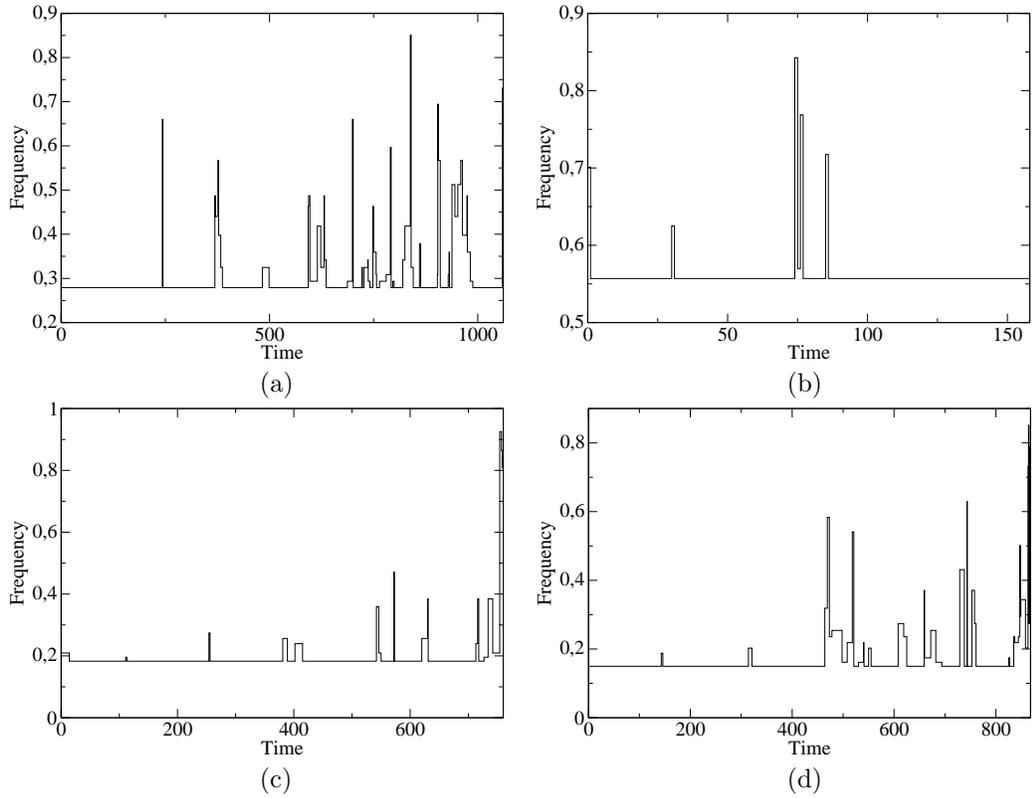

\begin{center}
 \begin{tabular}{ll}
  \scalebox{0.28}{\epsfig{file=dibujos/rss.eps,clip=}} &
  \scalebox{0.28}{\epsfig{file=dibujos/gmail.eps,clip=}} \\
  \hspace*{3.22cm} (a) & \hspace*{3.22cm} (b)\\
  \scalebox{0.28}{\epsfig{file=dibujos/atentado.eps,clip=}} &
  \scalebox{0.28}{\epsfig{file=dibujos/terror.eps,clip=}}\\
  \hspace*{3.22cm} (c) & \hspace*{3.22cm} (d)
 \end{tabular}
\caption[]{Fitting curves for some real world document streams; each
stream corresponds to blog posts from Blogalia that include the
following words: a) RSS; b) gmail c) {\em atentado} and d) {\em
atentado terrorista}}
\label{freales}
\end{center}
\end{figure}

Finally, we have studied the behavior of some streams obtained from
the real world. They have been obtained from the Blogalia,
weblog hosting site\footnote{\sf http://www.blogalia.com}, by doing a
database search on certain words: (a) {\bf rss}, (b) {\bf gmail},
(c) {\bf terrorismo}\footnote{terrorism} and (d) {\bf atentado
terrorista}\footnote{terrorist attack}. 
Figure \ref{freales} shows the fitting curve obtained by the EA with a population size 
of 200 individuals, a maximum number of 1000 iterations, a crossover rate of 40 \%,
a mutation rate of 5 \% and an automaton of 25 states. 
We can observe that the fit for {\em rss} (Figure \ref{freales} (a)) shows many peaks, although
they are mainly concentrated in the second part of the considered period of time.
This behavior can be expected for a stream of a topic on which the interest can be very
variable.  The curve for the  {\em gmail} stream (Figure \ref{freales} (b)) presents a clear burst on
two particular intervals of time approximately in the middle of the
considered period, which roughly corresponds to new waves of GMail
(the Google webmail system, at {\sf http://gmail.com}) invitations.
The two last fits (Figure \ref{freales} (c) and (d)) are devoted to two related topics,
terrorism and terrorist outrage,
and thus, we can observe that although the intensity of document arrivals is greater in (d),
because the topic is more general, the bursts in both streams appear approximately 
at the same times, as it can be expected.

\section{Dynamic Detection of Changes of Interest in Document Streams}

We have also investigated the dynamic detection of changes in the trends of documents
streams. Let us assume the fit for a stream has been found, new
documents arrive, and we want to detect possible changes in the trends 
of the corresponding topic. Clearly, the most accurate way of doing this is 
to apply again the algorithm for finding the best fit for the extended stream.
However, thinking about real-time applications for the web,
we have investigated ways to fit just one new document, without fitting the
whole stream until there is time to do it. 
This can be done by searching the minimum cost for the new gap $x$ according to
the last state of the previous fit. Thus, if the fit obtained
for the previous stream finished at state $q_i$, we look for the state $q_j$
which minimizes the cost
\[
 arg~min_{q_{j}} P(q_j|q_i,x) = arg~min_{q_{j}} (\tau(q_i, q_j) + \mbox{ln}~ {\alpha}_j e^{-{\alpha}_j x})
\]
In order to check if this ``local'' approximation is meaningful, we
have chosen one real world stream, the one tracking the term {\em
gmail}, and have performed some cuts on it. 
Table \ref{tfit_local} shows the results for some of these cuts. This mechanism,
which allows us to detect immediately changes in the trends of a stream,
has properly worked in all these cases, yielding the state that had
previously been found by the fitting algorithms.

\begin{table}[htbp]
\begin{center}
\begin{tabular}{|l|c|c|c|c|}
\hline\hline
Previous substream          & Arrival Time  & Old state & New state & Trend\\
\hline
$\cdots$ 38 38 39 41 49 49  & 52            & 12        &   0       & $\downarrow$\\
\hline
$\cdots$ 41 49 49 52 68 69  & 69            & 3         &   4       & $\uparrow$\\
\hline
$\cdots$ 88 89 90 90 91 92  & 95            & 0         &   0       & $\rightarrow$\\
\hline
\end{tabular}
\caption{Results of applying a local approximation to detect changes in the trend of
a stream.
}
\label{tfit_local}
\end{center}
\end{table}

Another improvement to quickly obtain a fit for an extension of a
previously fitted stream, 
is to use the previous fitting curve as a seed for the EA which
searches the fit of the extended stream.
To implement this mechanism, we have modified the way in which the EA creates the initial population.
A {\em seed individual} is created, its last gene extended with the new substream
of document dates. Then, the initial population is created by applying the mutation operator to
this seed individual, but in such a way that the last gene has a higher probability to undergo
mutation. Once the initial population has been created, the EA proceeds as before.

Experiments have shown that this mechanism can save a lot of execution time. 
Table \ref{tsemilla} shows the time required to fit the stream of Figure \ref{fartificial}(c)
if a part of it, whose length appears in the first column, has already been fitted.
We can observe that the time required is very small.

\begin{table}[htbp]
\begin{center}
\begin{tabular}{|c|c|c|c|}
\hline\hline
Subst. Len.  & New Subs. len.& T. w/out seed    & T. w/ seed \\
\hline
219900       & 100           & 3895.28     &  141.45 (79.09)\\
\hline
219000       & 1000          & 3895.28     &  144.75 (81.96) \\
\hline
220000       & 10000         & 3895.28     &  166.73 (79.32) \\
\hline
\end{tabular}
\caption{Time (seconds) spent without seed and using the previous fit as a seed for
some substream taken from the stream of Figure \ref{fartificial}(c).
The first column indicates the length of the  previously fitted substream,
and the second one the length of the stream of documents which has to be added to
the previous one. The third column presents the time needed to fit the
whole stream, while 
the last one presents the time spent to reach convergence by using the previous 
fit as seed, with a population size of 200 individuals, a crossover rate of 40 \% 
and a mutation rate of 10\%. The result for a population
size of 100 individuals appears in parentheses.
}
\label{tsemilla}
\end{center}
\end{table}

\section{Conclusions and future work}

In this paper, we have presented the design of a system devoted to the
dynamic detection of changes on 
the trends of the topics of a stream of documents, such as
newscasts, e-mails, IRC conversations, scientific journals or weblogs. 
It is based on modeling the assignment of frequencies to intervals
of document arrivals and obtaining an optimal fit to the data. 
We have designed an evolutionary algorithm to implement the model, 
what allows us to deal with very large sequences of documents in a 
reasonable time, obtaining fitting curves with a similar shape to 
those provided by classic dynamic algorithms. 
We have studied different issues of the model and
its implementation, such as the criterion to change to an interval
with a new frequency. We have found that penalizing the change of
an interval to another with different frequency, whether it is higher or 
lower, provides worse results that penalizing only those changes 
to an interval of higher frequency.
We have also tested different functions for this last type of penalization,
and found that the results depend weakly on this function. Because of this,
we prefer a function which normalizes the penalization with respect
of the number of frequency values considered,
in order to make the results independent of this parameter.

We have also designed a version of the evolutionary algorithm
which dramatically reduces the time required to find the optimal fit to a stream 
which is an extension of a previously fitted substream.
This version of the evolutionary algorithm uses the previous fit as
a seed to generate the initial population, which can quickly converge
if most of the stream has been previously fitted.
In this way, our system can be applied to dynamically model the document stream,
and thus detect changes on the trends of the corresponding topic in real time.
Besides, the fitting curves produced by the system  for a stream of 
documents can also be useful for other applications: 
the fit obtained for streams corresponding to different topics can help
to detect correlations between these topics, to study how a topic
affects others, etc.

Future lines of works will include:\begin{enumerate}
\item An exhaustive study of the evolutionary algorithm parameters, to
obtain better performance.
\item Study of correlation among document streams, to automatically
detect the occurrence of new topics composed of multi-word
concepts. This can also be helped by other techniques.
\item Optimization for real-time operation, including parallelization
of the algorithms. 
\item More extensive checking in many different document streams.
\item Characterization of document streams via its model.
\end{enumerate}

\bibliographystyle{unsrt}
\bibliography{topic,GA-general}

\end{document}